\documentclass[structabstract]{aa}  

\usepackage{graphicx}
\usepackage{lscape}
\usepackage{txfonts}
\usepackage{caption}
\usepackage{color}
\usepackage{grffile}
\usepackage{afterpage}
\usepackage{amsmath}

\newcommand{\at}{{ATLASGAL}}

\newcommand{\msol}{M$_{\odot}$}
\newcommand{\uchii}{{UC-H\,\scriptsize II}}

\begin{document}

  \title{The ATLASGAL survey -- distribution of cold dust in the Galactic plane}
   \subtitle{Combination with Planck data}
   \authorrunning{T. Csengeri et al.}
   \titlerunning{ATLASGAL  -- combination with Planck data}

 \author{T. Csengeri
          \inst{1}
          \and
          A. Weiss
          \inst{1}
          \and         
          F. Wyrowski 
         \inst{1}
          \and
          K. M. Menten 
          \inst{1}
          J. S. Urquhart
          \inst{1}
          \and
          S. Leurini
          \inst{1}
           \and
          F. Schuller
          \inst{2}
         \and
         H. Beuther
         \inst{3}
         \and 
         S. Bontemps
         \inst{4}
         \and 
         L. Bronfman
          \inst{5}
         \and 
         Th. Henning
          \inst{3}
         \and 
         N. Schneider
          \inst{4,6}
          }
  \institute{Max Planck Institute for Radioastronomy,
              Auf dem H\"ugel 69, 53121 Bonn, Germany\\
              \email{ctimea@mpifr-bonn.mpg.de}
         \and
             European Southern Observatory, Alonso de Cordova 3107, Vitacura, Santiago, Chile
         \and
             Max Planck Institute for Astronomy, K\"onigstuhl 17, 69117 Heidelberg, Germany
         \and
           OASU/LAB-UMR5804, CNRS, Universit\'e Bordeaux 1, 33270 Floirac, France    
         \and
           Departamento de Astronom\'{i}a, Universidad de Chile, Casilla 36-D, Santiago, Chile
         \and
           I. Physikalisches Institut, Universit\"at zu K\"oln, Z\"ulpicher Strasse 77, 50937 K\"oln, Germany}
           

 
  \abstract
  {
  Sensitive ground-based submillimeter surveys, such as ATLASGAL, provide
  a global view on the distribution of cold dense gas in the Galactic
  plane at up to {two-times} better angular-resolution compared to
   recent space-based surveys with Herschel.
  However, a drawback of  ground-based continuum observations is that they
  intrinsically filter emission, at angular scales larger than a fraction of the field-of-view
  of the array, when subtracting the sky noise in the data processing. 
  The lost information on the distribution of diffuse emission
  can be, however, recovered from 
   space-based, all-sky surveys with Planck.
   }
   {
   Here we aim to demonstrate how this information can be used
   to complement ground-based bolometer data and 
   present reprocessed maps of the APEX Telescope Large Area Survey of the Galaxy  
({\at}) survey. 
   }
   {
   We use the  
   maps at 353~GHz from the Planck/HFI instrument, 
   which performed a high sensitivity all-sky survey 
   at a frequency close to that of the APEX/LABOCA array, which is centred on 
   345~GHz. 
   Complementing the ground-based observations with information on larger angular 
   scales, the resulting maps
  reveal the distribution of cold dust in the inner
   Galaxy with a larger spatial dynamic range. 
   We visually describe the observed features and
  assess the global properties of 
    dust distribution.
   }
   {
   Adding information from large angular scales helps to better 
   identify the global properties of the cold Galactic interstellar medium.
   To illustrate this, we provide mass estimates from the dust towards
   the W43 star-forming region and estimate a column density contrast of at 
   least a factor of five between a low intensity halo and the star-forming ridge.
   We also show examples of elongated 
   structures extending over angular { scales} of $0.5^\circ$,
   which we refer to as thin giant filaments.
   Corresponding to $>30$ pc structures in projection at a distance of 3 kpc,
   these dust lanes are very extended and show large aspect ratios. 
   We assess the fraction of dense gas by determining the 
   contribution of the APEX/LABOCA 
    maps to the combined maps, and estimate 
   $2-5$\% for the dense gas fraction (corresponding to { $A_{\rm{v}}>$7 mag}) 
   on average in the Galactic plane.
  We also show probability distribution functions of the
  column density (N-PDF), which reveal the typically observed log-normal
  distribution for low column density and exhibit an {excess at high column densities}.
  As a reference for extragalactic studies,
  we show the line-of-sight integrated N-PDF of the inner Galaxy,
  and derive a contribution of this excess to the total column density of $\sim2.2$\%,
  corresponding to $N_{\rm H_2} = 2.92\times10^{22}$~cm$^{-2}$.
  Taking the total flux density observed in the maps,
  we provide an independent estimate of the mass of molecular
  gas in the inner Galaxy of $\sim1\times10^9$\,\msol, 
  which is consistent with previous estimates using CO emission. 
  From the mass and  dense gas fraction ($f_{\rm DG}$), 
  we estimate a Galactic SFR of 
  $\dot M = 1.3$\,\msol\,yr$^{-1}$.
    }
  {
  Retrieving the extended emission 
  helps to better identify massive giant filaments which 
  are elongated and confined structures.
  We show that the log-normal distribution of low column density gas is ubiquitous
  in the inner Galaxy. 
  While the distribution of diffuse gas is relatively homogenous in the
  inner Galaxy, the central molecular zone (CMZ)    
  stands out with a higher dense gas fraction
  despite its low star formation efficiency. 
  {
  Altogether our findings explain well the
  observed low star formation efficiency of the Milky Way by the low
  $f_{\rm DG}$ in the Galactic ISM. In contrast, 
  the high $f_{\rm DG}$ observed towards
  the CMZ, despite its low star formation activity, suggests 
  that, in that particular region of our Galaxy, high density gas is not the
  bottleneck for star formation.
  }
  }
  

   \maketitle
%

\section{Introduction}

   Large area surveys of the Galaxy in the (sub)millimeter regime
   are essential to study the distribution of  dust and star-forming gas.
   While space-based missions, such as \emph{Herschel}~
   \citep{Andre2010,Molinari2010,Motte2010} provide high sensitivity images
   at submillimeter wavelengths with unprecedented dynamic range, 
   ground-based facilities have the advantage of providing
   up to $3\times$ better angular resolution at a similar frequency. 
   In this context, the APEX Telescope Large Area Survey of the Galaxy  (ATLASGAL survey; \citealp{schuller2009}) provides one of the 
   most extensive ground-based
   mapping of the inner Galaxy at submillimeter wavelengths.

   Ground-based  bolometer arrays are, however, not well suited to   
   measure emission from angular scales larger than a fraction
   of the field-of-view of the array.
   This is because variations of the atmosphere 
   mimic emission from extended
   astronomical objects, and therefore this signal is removed when 
   subtracting the correlated noise from the maps in the data reduction process. 
   As a consequence, depending on the data acquisition method and the reduction,
   the final emission maps have limited sensitivity above certain angular
   scales. 
   With the advent of sensitive space-based missions, such as the 
   all-sky survey of Planck~\citep{Tauber2010},
   this missing information can now be retrieved.

   In particular, the Planck/High Frequency Instrument 
(HFI; \citealp{Lamarre2010}) surveyed the sky 
   at 353~GHz, i.e.\,at
   a similar frequency range as the APEX/Large APEX Bolometer Camera
   (LABOCA) at 345~GHz used
   for the \at\ survey.
   Here we present newly processed maps of 
   the \at\ survey that have been corrected for the loss of the filtered
   emission based on the Planck data.
   We describe the method in Sect.\,\ref{sec:obs}, then
   present and briefly describe the new data products in Sect.\,\ref{sec:res}.
   With this dataset, we investigate the large-scale properties of the Galactic cold dust
   in Sect.\,\ref{sec:stat}.


\section{Observations and data processing}\label{sec:obs}
\subsection{APEX/LABOCA}
The \at\ survey imaged 
in total $\sim420$~sq.\,degree of the inner Galactic plane at 870~$\mu$m 
(centred on 345~GHz) with the 
LABOCA camera \citep{siringo2009} on the APEX Telescope \citep{gusten2006}
at a 19$\rlap{.}{''}$2 spatial resolution.
The highest sensitivity part of the \at\ survey \citep{schuller2009} covers the Galactic plane between Galactic longitude,
$-60^\circ \le\ell \le +60^\circ$, and Galactic 
latitude, $-1.5^\circ \le b \le +1.5^\circ$, with an rms noise of
$\sim70-90$~mJy/beam.
The survey was then extended towards Galactic longitude $-80^\circ \le \ell \le -60^\circ$ and Galactic latitude $-2^\circ \le b\le+1.0^\circ$ with an rms noise of
$\sim110$~mJy/beam (see also~\citealp{Csengeri2014}).

The observing strategy and the main steps of the data reduction procedure are 
described in detail by \citet{schuller2009}. The data were reduced with the {\sl BoA} 
software \citep[]{schuller2012}\footnote{http://www.eso.org/sci/activities/apexsv/labocasv.html}.
As a result of removing the correlated noise  in the time series of bolometer signals, the
information of emission from extended 
astronomical objects is lost. This can be partially recovered 
through an iterative data reduction process (see also~\citealp{Belloche2011}).  
 The data reduction of the \at\ survey has been optimized 
 for compact sources, and convergence was reached 
 after $15$ iterations. The maps are sensitive to emission
 with angular scales up to $2\rlap{.}'5$. 
 The absolute flux uncertainty is estimated to be less 
than $\sim15\%$ \citep{schuller2009}. 
The catalogue optimized to extract the properties of the embedded 
compact sources is provided in \citet{Csengeri2014}, 
while a catalogue of the more extended sources is 
presented in 
Contreras et al. (2013; see also \citealt{Urquhart2014}).
\subsection{Planck/HFI}
Planck performed an all-sky survey between 
30 and 857~GHz~\citep{PlanckI}, 
mainly focusing on measuring the 
structure of the cosmic microwave background radiation (CMBR). For this purpose, it is mandatory that all foreground emission is carefully measured, modelled, and subtracted. To this end, the differing wavelength dependences of the foreground emission's components (mainly, synchrotron, thermal free-free, and dust emission) are fully characterised by all sky maps made in a total of nine frequencies between 30 and 857 GHz. Of relevance here is that the HFI instrument
(\citealp{Lamarre2010, Planck2013_IX})
provides high sensitivity maps of the dust at 353~GHz
with an angular resolution of $4\rlap{.}'8$.

In the bright region of the Galactic plane, it is reasonable to assume that 
the 353~GHz emission is dominated by the interstellar dust, and
 CMBR, { free-free emission}, cosmic infrared background (CIB), and zodiacal dust emission {have}  minor
(${<5}$\%)
contamination~(see e.g.\,\citealp{PlanckXII, Planck2013_zodiacal_light, Planck2014_XIV, Planck2015_GBcomponents} and references therein). 
As a result of  similar centre frequencies, overlapping bandpasses and sensitivity 
at nearly matching angular scales,
this dataset 
 therefore well complements 
the \at\ survey  to
recover emission from larger spatial 
scales. 
For this purpose, we used the publicly
available Release 1 maps of the Planck mission~\citep{PlanckI}.

\subsection{{ Colour corrections for LABOCA and the HFI}\label{Sect-CC}}
\begin{figure}[!htpb]
\centering
\includegraphics[width=0.9\linewidth, angle=0]{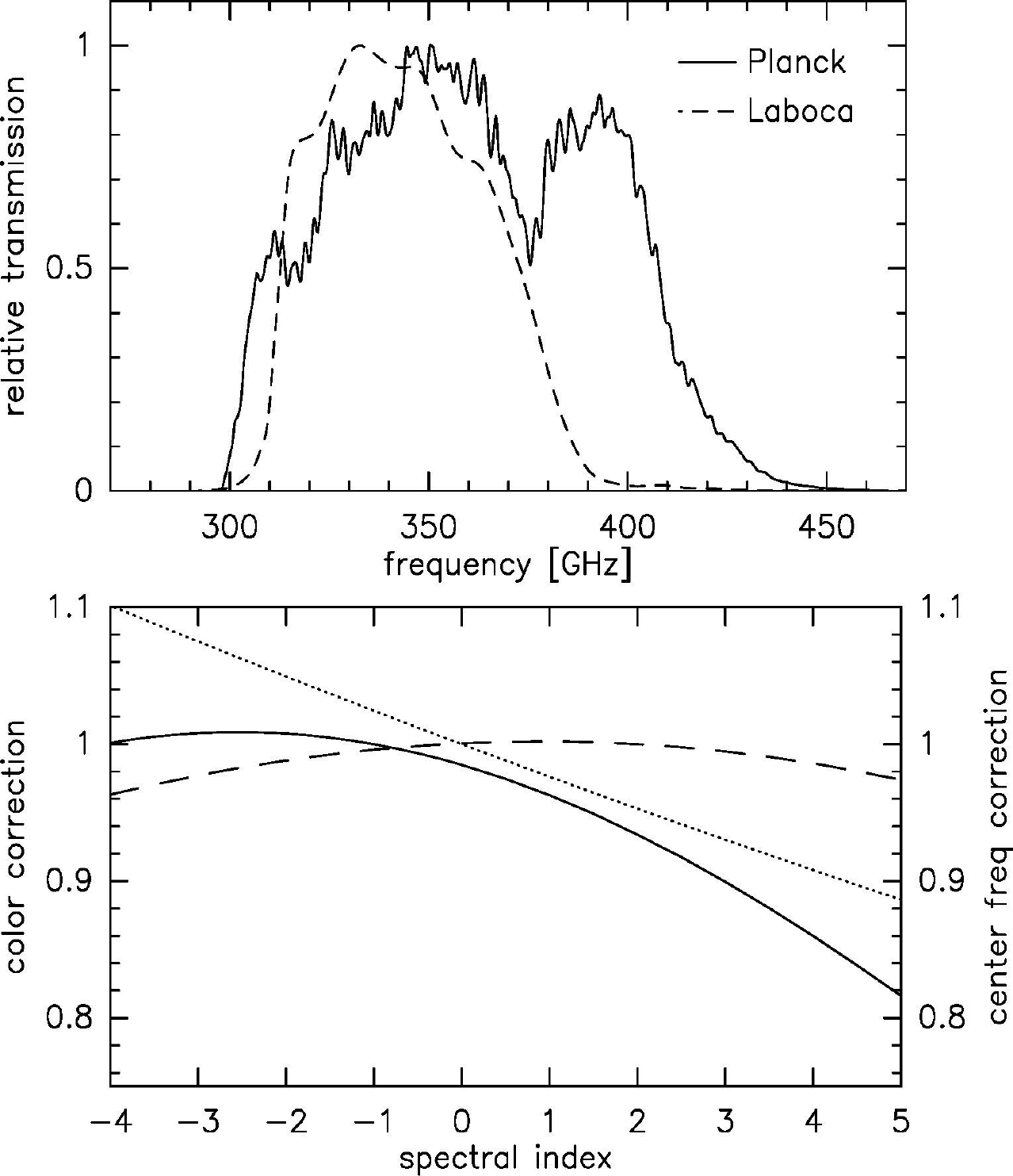}
\caption{{\it Top}: Relative spectral response (passband) for Planck's HFI-353GHz
channel (solid line, \citealp{Planck_explanatory_supplement}) and LABOCA (dashed line, \citealp{siringo2009}).
{\it Bottom}: {Colour corrections for the HFI-353 (solid line) and LABOCA (dashed line) as a function of the spectral
index of the assumed source spectrum. The dotted line shows the conversion factor to scale the HFI-353 intensities 
to the nominal centre frequency of LABOCA.}
}
\label{fig:color_correction}
\end{figure}

{ Since the calibration and  spectral shape of the passbands of 
LABOCA and Planck's HFI-353~GHz channel 
are different (see Fig.\,\ref{fig:color_correction} upper panel), colour corrections need 
to be applied to combine both datasets. Two corrections have to be taken 
into account: 
a) the difference of the nominal centre frequency for both intstruments; and b) colour corrections 
to the intensity calibration. Both depend on the 
intrinsic source spectrum and we here assume that the spectrum across the 
passbands, $S(\nu)$, can be approximated by a power law with $S(\nu) = S_{{\nu_0}}\,(\frac{\nu}{\nu_0})^{\alpha}$, 
where $\nu_0$ is the nominal centre frequency and $\alpha$ is the spectral index of 
the source spectrum.  The nominal centre frequencies for the HFI-353 and LABOCA are 353 and 345\,GHz, respectively.
The centre frequency correction from the HFI-353 to LABOCA is therefore given by $F(\alpha) = (345/353)^{\alpha}$.
The intensity calibration for the HFI-353 is based on the IRAS convention 
(I$_{\nu}\,\cdot\nu$=const), while the LABOCA calibration is based on 
planet spectra (I$_{\nu}\propto\nu^2$). Colour corrections to other source spectral indices
can be computed by 
\begin{equation}
C_{\rm HFI-353}(\alpha) = \frac{\int R(\nu)(\nu/353)^{-1}d\nu}{\int R(\nu)(\nu/353)^{\alpha}d\nu}
\end{equation}
and 
\begin{equation}
C_{\rm Laboca}(\alpha) = \frac{\int R'(\nu)(\nu/345)^2d\nu}{\int R'(\nu)(\nu/345)^{\alpha}d\nu}
,\end{equation}
where $R(\nu)$ and
$R'(\nu)$ are the HFI-353 and LABOCA passbands, respectively.
These corrections are shown as a function of the source spectral index in the bottom panel 
of Fig.\,\ref{fig:color_correction}. \\
The average spectrum of the disk of the Milky Way is well described
by a modified black-body spectrum with $\beta=1.8$ \citep{Planck2011_dust_nearby, Planck2011_dust, Planck2013_dust_model}. We therefore compute the corrections using a 
spectral index of $\alpha=3.8,$ which yields $F = 0.917$, $C_{\rm HFI-353} = 0.868, $ and $C_{\rm Laboca}= 0.988$.
The HFI-353~GHz data is provided in K$_{\rm CMB}$ brightness temperature units. Conversion to MJy\,sr$^{-1}$ (IRAS convention)
is provided by a factor of $U$=287.450\,MJy\,sr$^{-1}\,{\rm K}^{-1}_{\rm CMB}$ \citep{Planck2013_IX}.
Flux conversions to a common reference frequency at 345\,GHz and source spectral index of $\alpha=3.8$ thus include
\begin{equation}
S_{{\rm HFI},345,3.8} = F\,\cdot\,C_{\rm HFI-353}\,\cdot\,U\,\cdot\,T_{\rm CMB}\,\,[\rm {MJy\,sr}^{-1}]
\end{equation}
and
\begin{equation}
S_{{\rm Laboca},345,3.8} = C_{\rm Laboca}\,\cdot\,S_{\rm Laboca}\,\,[{\rm Jy\,beam}^{-1}].
\end{equation}
}

%
\begin{figure}[!htpb]
\centering
\includegraphics[width=0.9\linewidth, angle=0]{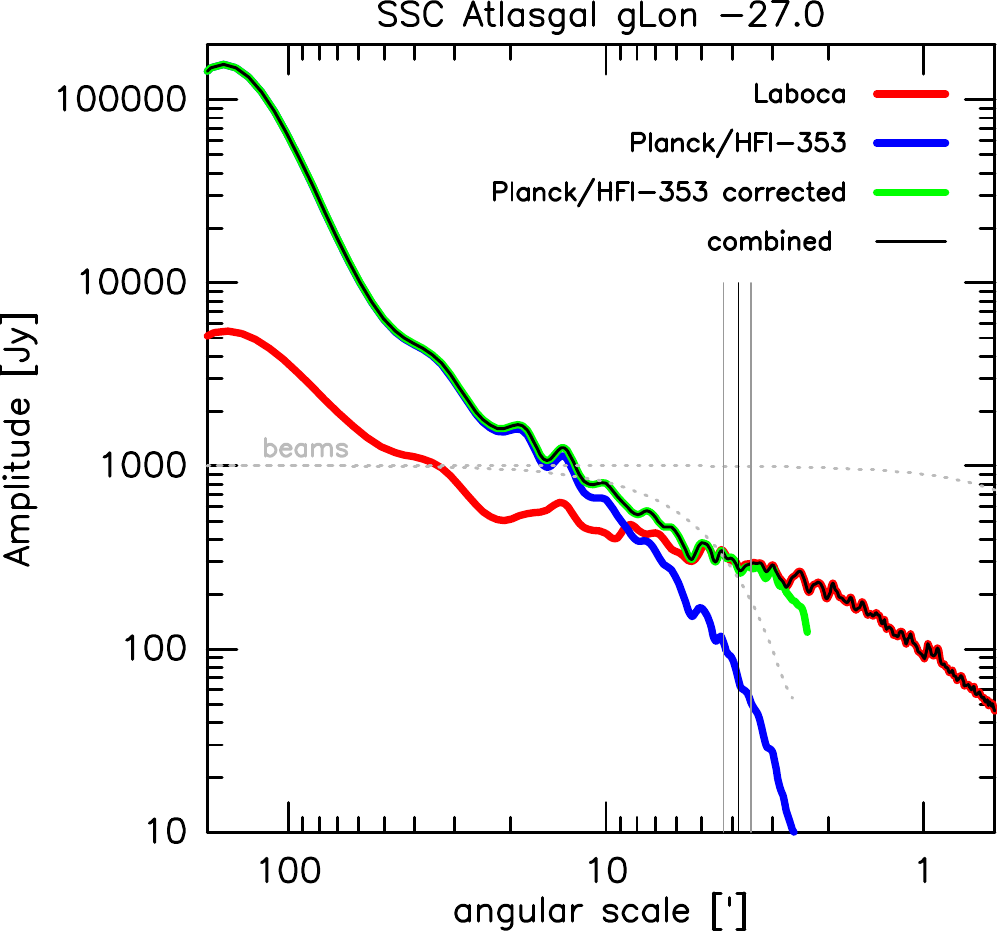}
\caption{
Amplitude versus spatial scales for the APEX/LABOCA 
(red), original Planck/HFI (blue), corrected Planck/HFI (green) and the combined
dataset (black). The vertical lines indicate the centre and the effective width
of the Butterworth filter used to combine both visibility sets. The dashed
grey lines indicate the visibility amplitudes of the normalised 
LABOCA and Planck/HFI beams (scaled to a peak amplitude of 1000 for better
visualisation).
}
\label{fig:fft}
\end{figure}

\subsection{Combination of LABOCA and HFI}

The procedure to correct for the filtered emission maps of ground-based bolometers is very
similar to that used to account for the missing short-spacing information in interferometric
observations. With the aim of recovering the larger scale emission, we
therefore combined the APEX/LABOCA with the Planck/HFI data, where the
latter provides the so-called short-spacing information.

We combined the two datasets in the {Fourier domain ($uv$ plane thereafter)}
with the method
described in detail in \citet{Weiss2001}. In brief, the combination
replaces the central part of the $uv$ plane of the APEX/LABOCA data
(which is affected by the filtering) with the appropriate values
calculated from the Planck/HFI data. The method relies on the accuracy
of the absolute calibration of the two maps and only requires
knowledge of the shape of the telescope beam for both datasets. The
LABOCA beam is approximated by a circular Gaussian with a FWHM of
$19\rlap{.}''2$ \citep{siringo2009}. The effective Planck/HFI beam shape 
is a function of the position on sky \citep{PlanckXII}. To determine
the beam variation across the \at\ survey area, we fit a 2D Gaussian
to the effective beams retrieved from the Planck data archive at different 
Galactic longitudes. This test shows that the beam's major and minor FWHM
do not change significantly across the region of interest and we 
used the mean effective beam parameters for the Planck/HFI with 
FWHM of $5\rlap{.}'19\times4\rlap{.}'52$  \citep{PlanckXII}. The position angle, however, 
rotates from ${\rm PA}=-12^{\circ}$ at $\ell$=280$^{\circ}$ to 
${\rm PA}=60^{\circ}$ at $\ell$=330$^{\circ}$ and stays roughly 
constant out to $\ell$=60$^{\circ}$. In the combination we therefore used a
variable position angle with values determined from a linear interpolation
of the effective beams fits derived at sampling points spaced by 10$^{\circ}$ in
Galactic longitude between $\ell=280-330^{\circ}$ and ${\rm PA}=60^{\circ}$ for
larger Galactic longitudes.\\

\begin{figure*}[!tpb]
\centering
\includegraphics[width=17cm, angle=0]{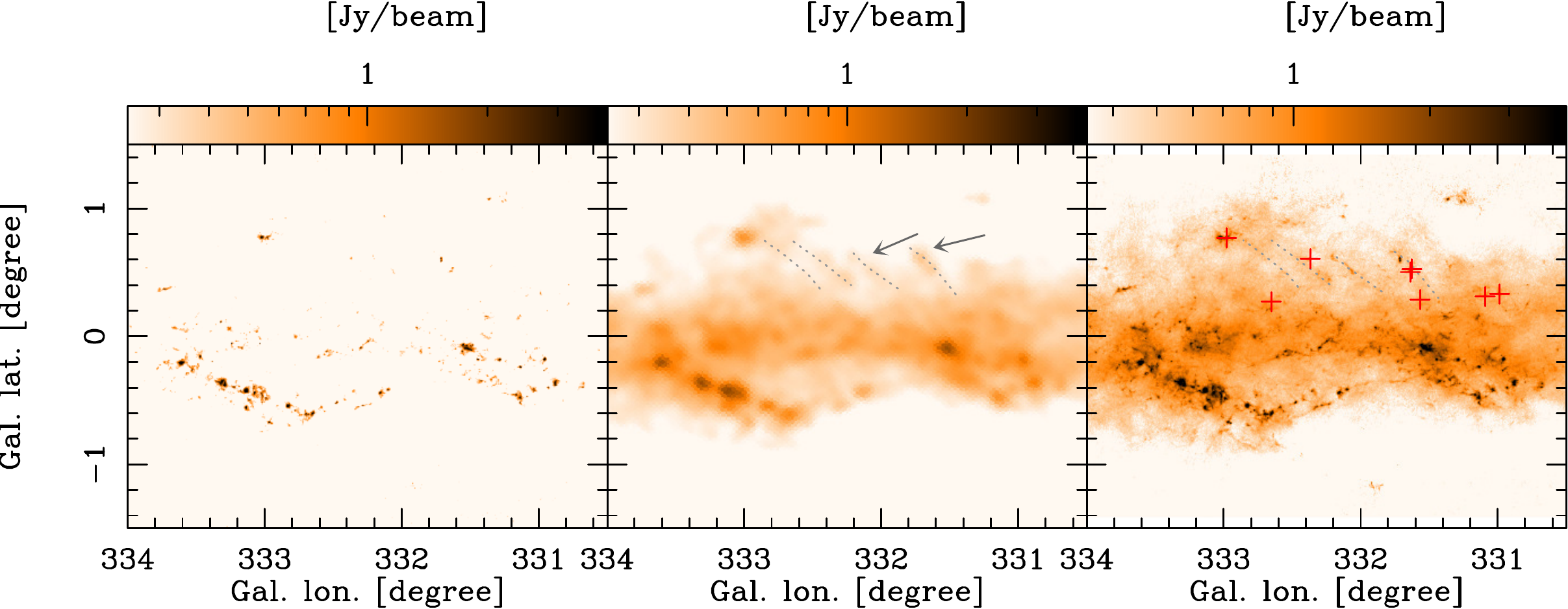}
\caption{
An example field showing the original APEX/LABOCA maps from the \at\
survey ({\bf left}), the corresponding tile from Planck/HFI ({\bf middle}) converted to 870\,$\mu$m, 
and the combined map ({\bf right}). The intensity scale is identical on the left and right panels
going from 0.35~Jy/beam to 4~Jy/beam on logarithmic scale.
Four arc-shaped, equidistant features bended to the Galactic plane
are indicated in grey dotted lines and with arrows.
Red crosses indicate the position of the velocity measurements from
\citet{Wienen2015}.
}
\label{fig:example}
\end{figure*}

{ First, we scale the Planck/HFI and the LABOCA intensities to a common reference 
frequency at 345\,GHz and source spectral index of $\alpha=3.8$ as described 
in Sect.\,\ref{Sect-CC}.} Then
we re-grid the Planck/HFI map to the same spatial grid as the LABOCA
map and convert both maps to a Jy\,pixel$^{-1}$ intensity scale. Next, both
maps and  beams are Fourier transformed to the $uv$ domain, the Planck/HFI
visibility\footnote{{ The term visibility refers to the complex numbers of the Fourier 
transform of surface brightness distribution.}} 
amplitudes are divided by the amplitudes of the Planck/HFI
beam (deconvolution of the Planck/HFI data) and the result is finally
multiplied by the amplitudes of the LABOCA beam (convolution of
the Planck/HFI data with the LABOCA beam). In these steps, the
amplitudes of both beams are normalised to 1 in the $uv$ domain. \\

After these operations the amplitudes of the Planck/HFI and APEX/LABOCA can
be compared to each other to determine which part of the $uv$ plane of
the APEX/LABOCA is affected by filtering and needs to be replaced by 
the Planck/HFI data (the only free parameter in the combination). 
An example of the generated amplitudes versus the spatial frequencies based on 
a $3\times3$ degree field is shown in Fig.\,\ref{fig:fft}. The figure 
demonstrates that for spatial scales larger than $\sim\,5'$ the APEX/LABOCA 
amplitudes fall systematically below the corrected Planck/HFI values due to the spatial 
filtering. Likewise, for scales smaller than $\sim\,2\rlap{.}'5$ the corrected 
Planck/HFI visibility amplitudes deviate from the LABOCA values because of our simplistic 
Gaussian approximation of the Planck/HFI beam shape. We use a Butterworth weighting 
function centred at a scale of $3\rlap{.}'8$ to combine the corrected Planck/HFI and the 
LABOCA visibility amplitudes. The width of the weighting function was set such that
for scales $>4\rlap{.}'3$ visibilities are purely based on Planck/HFI, and for scales $<3\rlap{.}'5$ 
visibilities are purely based on LABOCA. The weighting function is indicated by the 
vertical lines in Fig.\,\ref{fig:fft}. Finally, the combined
data is transformed back to the image domain and converted to the Jy/beam flux density
scale.\\

To test the relative calibration between the datasets and the critical Planck/HFI
beam model, we  computed the relative difference between the amplitudes
of the LABOCA and corrected Planck/HFI visibilities for spatial scales between
$2'.5-5'$. We find a good agreement between both datasets 
with an rms of { 15.5\%} across the survey.
{The relative small angular scale of $3\rlap{.}'8$ chosen for the combination
implies large correction factors for the Planck/HFI visibility amplitudes (up to 8.2). Because of
the much higher signal-to-noise ratio of the Planck/HFI data compared to the LABOCA data,
however, this has a negligible effect on the noise properties of the combined map, which is dominated
by the LABOCA noise. In the few outer regions of the survey where no significant emission is seen in the Planck/HFI data 
the noise level of the combined map is only $\sim5\%$ higher than in the \at\ maps.}\\

In practice, the combination is performed with the MIRIAD software package \citep{Miriad}
on $3.0\times3.0$ degree subfields of the \at\ survey. To avoid 
Fourier-transform artifacts from strong emission at the edge of 
these tiles, we  multiplied the LABOCA and the re-grided Planck/HFI data 
prior to transformation to the $uv$ domain with a 2D Butterworth filter. This
effectively reduces the size of each zero-spacing corrected tile to $2.8\times2.8$ 
degrees. To compensate for this edge effect along the Galactic plane, the individual
tiles are spaced by 1.5 degree in Galactic longitude so that no information is lost
except for the 0.1 degree broad edges in Galactic latitude.
The combined data are publicly available at {\tt http://atlasgal.mpifr-bonn.mpg.de/}.

To check the consistency of the method, we compared the total flux density
in the original Planck/HFI maps and the combined maps, which were
found to agree to better than $5$~\%. This is 
less than the calibration uncertainty of the LABOCA data.


\begin{figure*}[!htpb]
\centering
\includegraphics[width=18cm, angle=0]{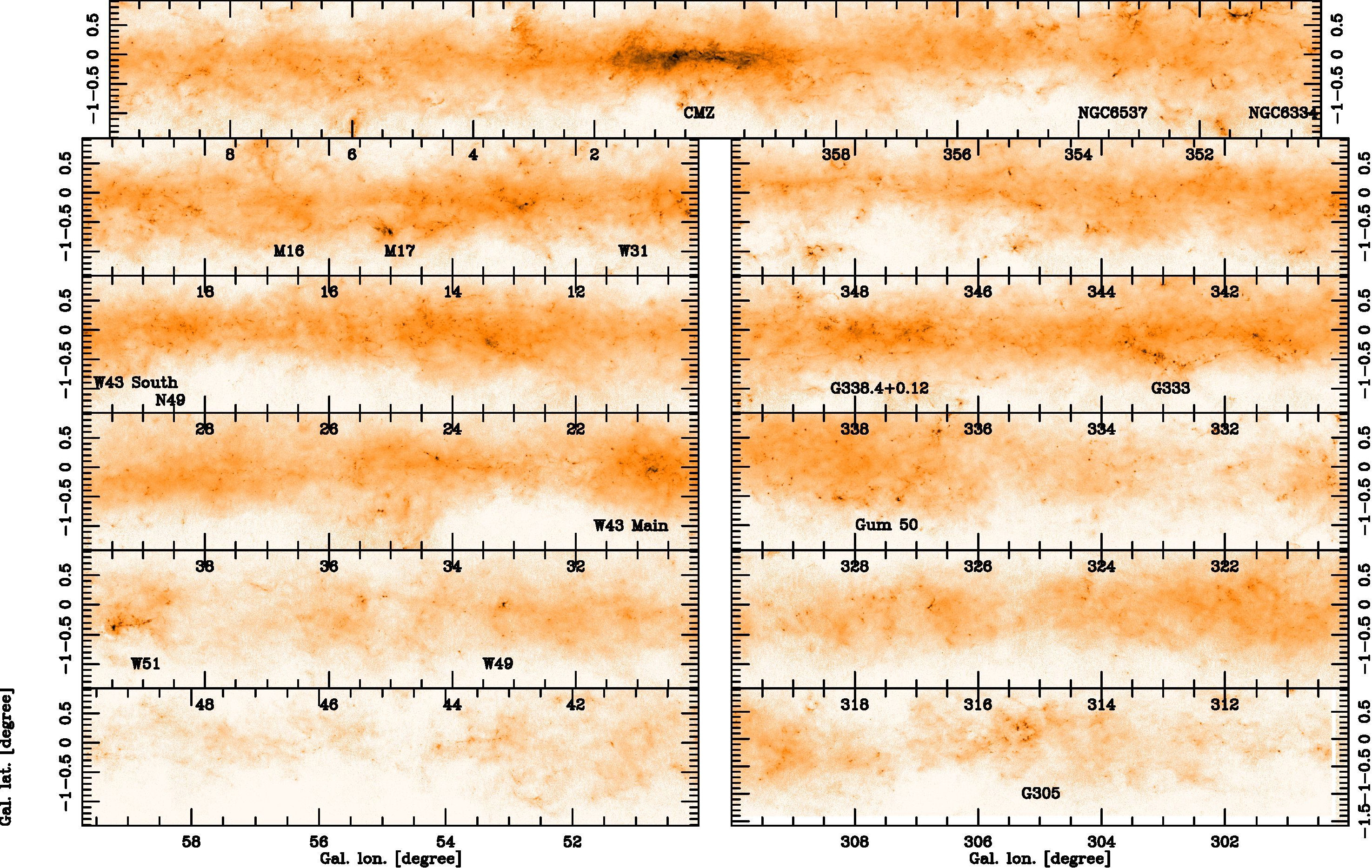}
\caption{
Overview of the Galactic dust emission with the Planck+APEX combined
maps at 870~$\mu$m. Colour scale goes from 0.15 to 5~Jy/beam on logarithmic scale.
GMCs and cloud complexes are labelled based on \citet{Wienen2015}.
}
\label{fig:all}
\end{figure*}

\section{Results}\label{sec:res}
\subsection{Large-scale structures in cold dust}

In Fig. 3, we show the maps, before and after the combination procedure,
 of one of the
most spectacular regions of the \at\ survey, 3 degrees wide centred at $\ell\sim332^\circ$. 
The extended emission clearly adds new information to
the original LABOCA maps, and the large-scale structures 
become more prominent and seemingly connected in the 2D projection of the dust.

In particular, we visually identify  
elongated diffuse structures bending to the Galactic plane. These arm-like
structures, extending over almost $0.5-1^\circ$, are visible in several
regions along the inner Galaxy. Prominent examples
are shown in Fig.\,\ref{fig:example} with several, equidistant dust lanes
that are parallel to each other and are distributed over a spatial extent of $\sim2^\circ$.
Pointed spectral line observations of high density tracers towards the
embedded compact sources by \citet{Wienen2015}
suggest a dispersion in $v_{\rm lsr}$ between $-57$ to $-42$~km/s,
corresponding to kinematic distance estimates of $3-3.5$~kpc.
Located at these distances their extension is $\sim30-60$~pc large.
These features are qualitatively similar to the results of hydrodynamic simulations of our 
Galaxy (see for example \citealp{Dobbs2011,Dobbs2015}), which produce equidistant
streams of inter-arm material produced by shears between spiral arms. 
A more detailed comparison to numerical models is, however, needed to interpret
these features observed in the large-scale structure of the Galactic cold gas.


Figure\,\ref{fig:all} shows an overview of the APEX/LABOCA and the
Planck/HFI combined maps {(in the following, we simply refer to them as combined maps)} 
over the Galactic plane in the highest sensitivity
part of the survey covering $|\ell|<60^\circ$. This complete view of the cold dust in the inner Galaxy clearly reveals the confined dust lane of our Galaxy, which is brightest between at $\ell \lesssim 40^\circ$ and $\ell\gtrsim312^\circ$. Outside
these longitude ranges the intensity of the emission along the plane gets significantly weaker.

\begin{figure}[!htpb]
\centering
\includegraphics[width=3.8cm, angle=270]{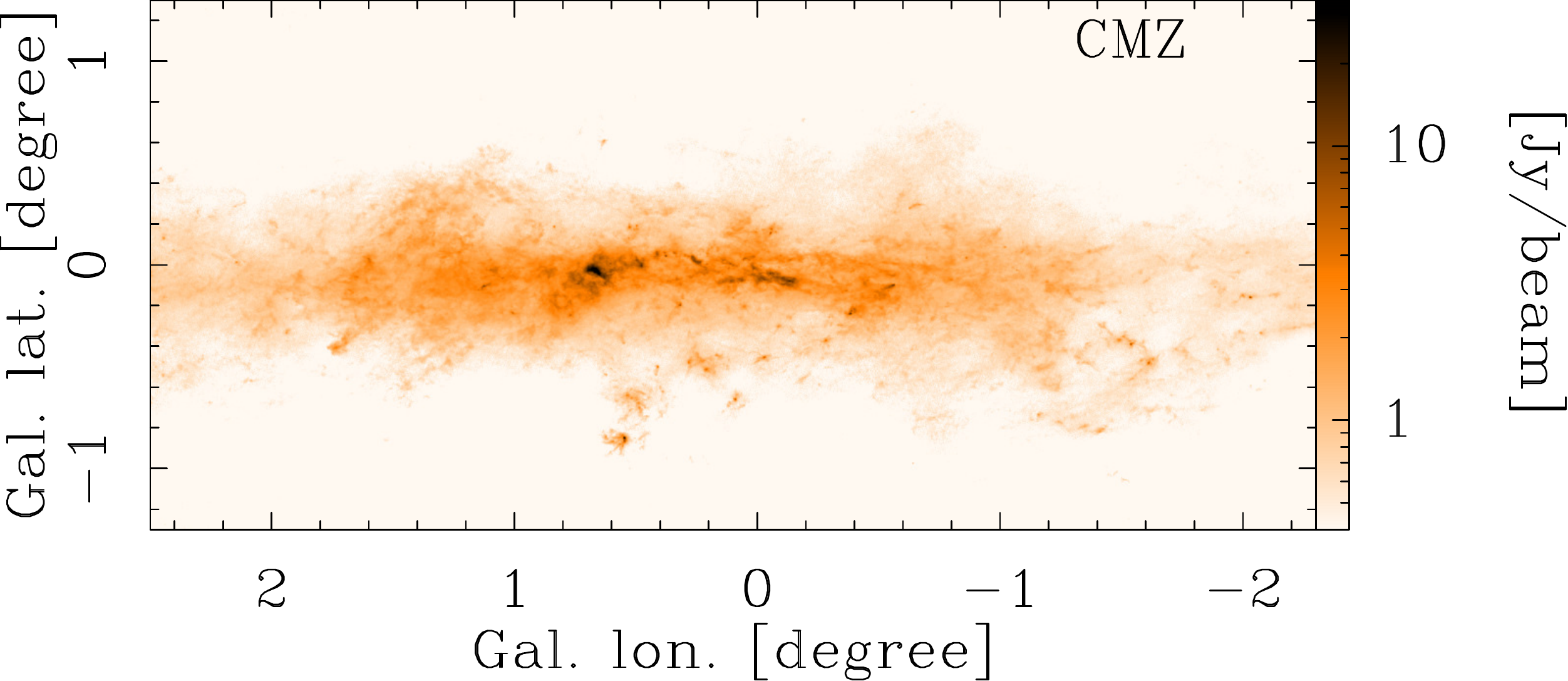}
\caption{
CMZ shown in the combined maps. 
The intensity scale goes from 0.4 to 34~Jy/beam on logarithmic scale.
}
\label{fig:cmz}
\end{figure}

The brightest and most prominent emission is
observed towards the Galactic centre and the central molecular zone (CMZ) extending
over several degrees (Fig.\,\ref{fig:cmz}). 
The other most prominent large-scale structures are 
associated with the brightest known star-forming 
giant molecular clouds (GMCs). 
These were also found to be coherent structures in the velocity domain 
 by \citet{Wienen2015}. 

\begin{figure}[!htpb]
\centering
\includegraphics[width=8.5cm, angle=0]{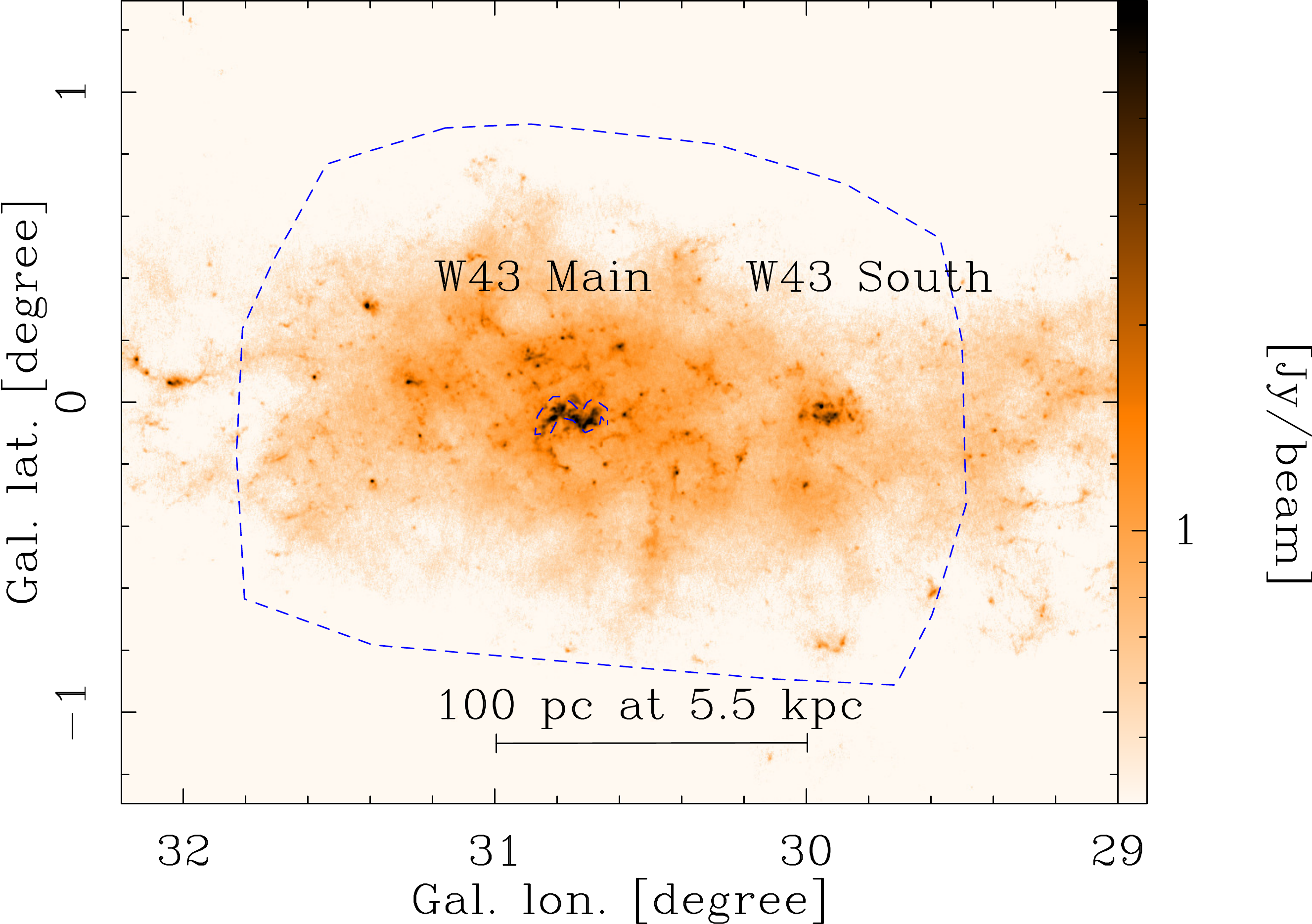}
\caption{
W43 GMC shown in the combined maps. 
The intensity scale goes from  0.4 to 6 Jy/beam on a logarithmic scale.
Dashed lines indicate the area where the emission has been summed up
for mass and column density estimates (see text for details).
}
\label{fig:w43}
\end{figure}

As an example, in Fig.\,\ref{fig:w43} we show the W43 GMC~\citep{Quang2011, Motte2014}, which
seems to be surrounded by a large, extended halo 
{ over an area of $\sim$\,$200\times160$\,pc seen in projection, and}
showing lower intensity diffuse emission.
To estimate the total mass towards this region, including the low intensity diffuse
gas, we sum up the emission in the area outlined in Fig.\,\ref{fig:w43}. 
We also use the formulae and the same assumptions as in \citet{schuller2009} and \citet{Csengeri2014}, assuming optically thin emission of dust at 870~$\mu$m
\begin{equation}
M [M_\odot]=\frac{S_\nu\,R\,d^2}{B_\nu(T_d)\,\kappa_{\nu}} 
\simeq 6.33 \times S_\nu \times \Big( \frac{d}{[\rm kpc]} \Big)^2
,\end{equation}
where {$S_\nu$ is the integrated flux density over the selected area}, and 
$d$ corresponds to the heliocentric distance.
To be consistent with our previous studies (e.g. \citealp{Csengeri2014}),
we use a gas-to-dust mass ratio (R) of 100 and  $\kappa_{\nu}=1.85$ cm$^2$~g$^{-1}$
for $\nu=345$\,GHz from Table 1, Col.\,5 of \citet{OH1994}. 
The numerical constant is obtained using 
a typical dust temperature dominated by the interstellar radiation
field of $T_{\rm d}=18$~K \citep{Bernard2010}.
Here we adopt a distance of 5.5~kpc 
based on maser parallax measurements towards W43 \citep{Zhang2014}.
This gives a total mass estimate from the dust of $\sim$\,${ 1.1\times10^7}$~\msol,
which is in a good agreement
with the $8\times10^6$~\msol\ estimate of \citet{Motte2014} based on molecular gas.

To calculate the H$_2$ column density contrast between the diffuse halo around the GMC
and the star-forming sites, we use the following expression:
\begin{equation}\label{eq:nh2}
N({\rm H_2})=\frac{F_\nu\,R}{B_\nu(T_d)\,\Omega\,\kappa_{\nu}\,\mu_{\rm H_2}\,m_{\rm H}}
\simeq 2.945\times10^{22}\times\frac{F_\nu}{[\rm Jy\,beam^{-1}]}\,\rm{[cm^{-2}]
}\end{equation}
where {$F_\nu$ is the flux density}, $\Omega$ is the solid angle
of the beam calculated by $\Omega = 1.13\times \Theta^2$, where 
$\Theta$ is the beam FWHM. 
The $\mu_{\rm H_2}$ is 
 the mean molecular weight of the interstellar medium
with respect to hydrogen molecules, which is equal to 2.8 \citep{Kauffmann2008},
and $m_{\rm H}$ is the mass of an hydrogen atom. The numerical
constant is obtained for a centre frequency of 345~GHz, 
$T_{\rm d}=18$~K, and beam size smoothed to 21\arcsec. 

We find a mean
H$_2$ column density of ${1.71\pm0.88\times10^{22}}$~cm$^{-2}$ in the 
{ larger environment of the W43 complex. This includes the halo 
and excludes the most active site of star formation associated
with the Z-shaped region of W43~Main, also described by \citet{Bally2010}
and \citet{Louvet2014}, shown in a smaller polygon in Fig.\,\ref{fig:w43}.}
Here we measure a mean H$_2$ column density of 
$7.45\pm5.51\times10^{22}$~cm$^{-2}$,
which implies an increase of column density contrast of $>5$ 
between the low intensity halo and the star-forming massive ridges
{ going from $\sim$\,$200$ pc to $\sim$\,$10-20$\,pc} scales
(Fig.\,\ref{fig:w43}).



\subsection{Thin giant filaments}

\begin{figure}[!htpb]
\centering
\includegraphics[width=0.9\linewidth, angle=00]{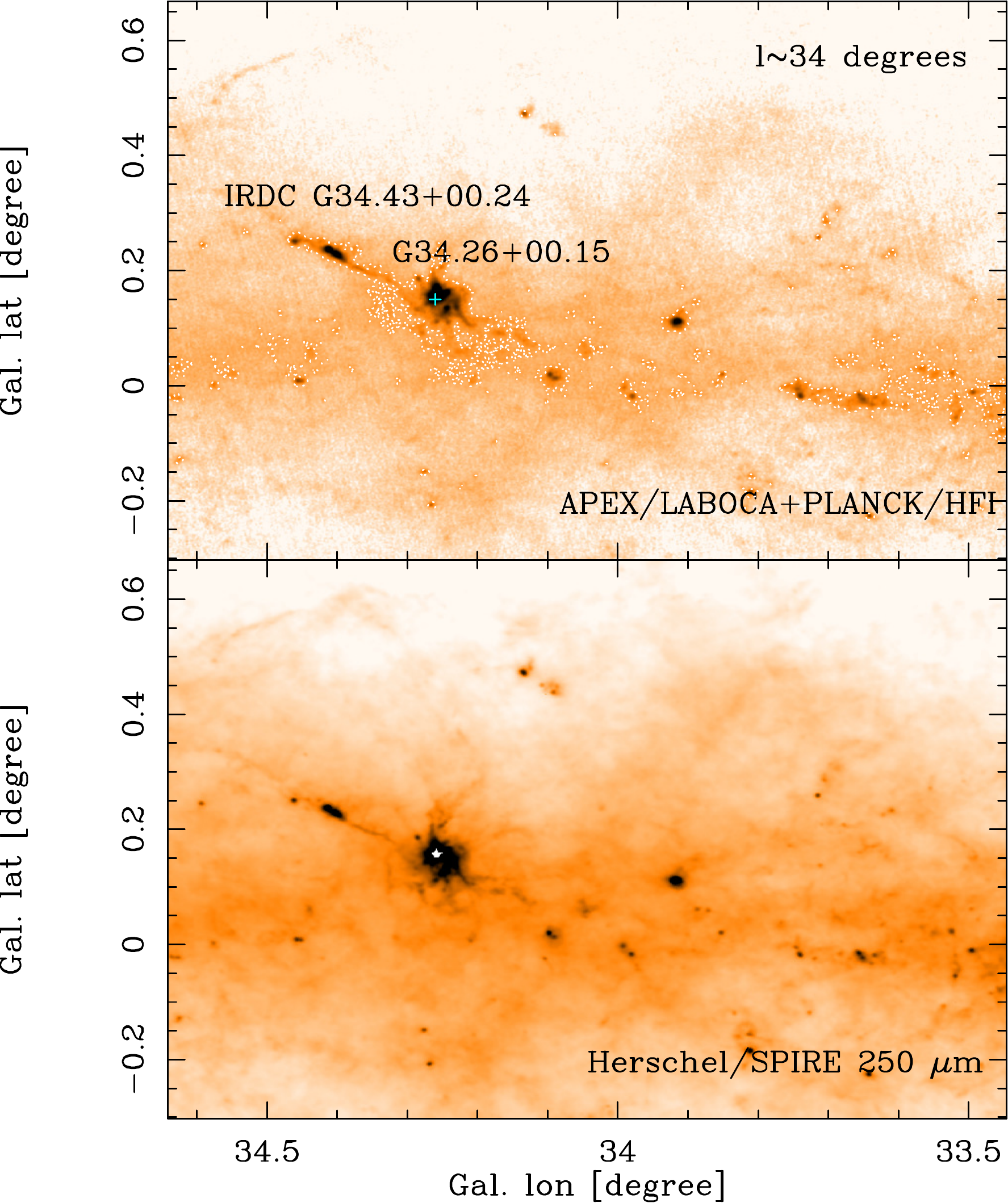}
\caption{{ {Top:}}  APEX/LABOCA and Planck/HFI 
combined map of the IRDC~IRDC G34.43+00.24.  
Grey contour shows an H$_2$ column 
density limit of $2.1\times10^{22}$~cm$^{-2}$.
A cross indicates the position of the 
{\uchii} region of G34.26+0.15.
{ {Bottom:}} \emph{Herschel}/250~$\mu$m map
of the same region from the Hi-GAL survey \citep{Molinari2010}.
{We show level three data products obtained from the publicly available Herschel Science Archive.}}
\label{fig:filaments}
\end{figure}

By revealing the structure of the 
lower density material connecting the highest density filament-fragments
revealed by ATLASGAL, 
the combined maps allow us to
visually identify
thin, but spatially { elongated} filamentary structures. An example
is shown in Fig.\,\ref{fig:filaments}, revealing the large-scale environment of 
the IRDC G34.43+00.24~\citep{Garay2004}.
Since ATLASGAL has a similar angular resolution
as the \emph{Herschel}/SPIRE 250~$\mu$m band,
 the strength of the survey lies in its sensitivity to cold dust.
To illustrate this, in Fig.\,\ref{fig:filaments} we also show 
the SPIRE~250~$\mu$m band maps observed 
within the Hi-GAL Herschel Key Programme~\citep{Molinari2010},
{which has similar angular resolution as the combined maps}.
Since the highest density central regions of these filaments are colder than the 
diffuse surroundings, they show a higher contrast in the 
ATLASGAL maps, as for instance in the  \emph{Herschel}/SPIRE 250~$\mu$m image.

Typical examples of these kinds of confined filamentary structures are the Snake (G11.11-0.12) \citep{Pillai2006, Henning2010}
and the Nessie filament~\citep{Jackson2010,Goodman2014}, 
which have been studied in great detail.
These structures have been characterised as giant molecular
filaments (e.g.\,\citealp{Li2013}; \citealp{Ragan2014})
and massive molecular filaments \citep{Battersby2012},
and based on their dust emission have also been identified by 
\citet{Wang2015}. 
The filaments we report here
typically extend over $\gtrsim0.5^\circ$ seen in projection, 
and are elongated structures. 
Taking a typical distance of
3--6~kpc, such structures have a linear extension larger
than  27--54~pc.
The presented new maps of the ATLASGAL
survey contain a handful of these visually identified structures,
however, it is beyond the scope of this paper to discuss them in detail.

\begin{figure}[!htpb]
\centering
\includegraphics[width=7.cm, angle=90]{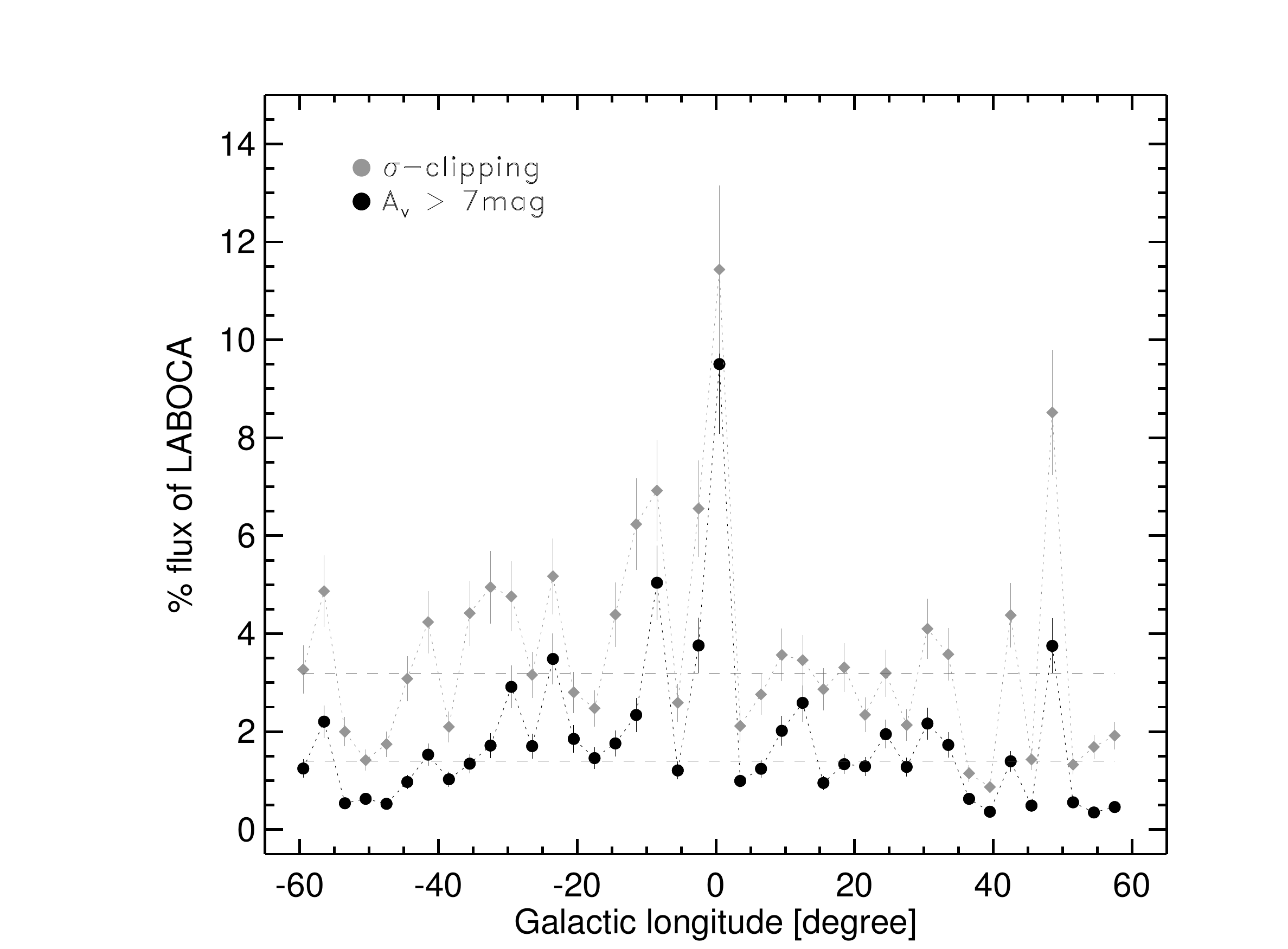}
\caption{
Percentage of the total flux in the APEX/LABOCA maps
compared to the combined maps. Black dots correspond to
the total flux estimated from all positive pixel values, while
grey dots are estimated from summing up all pixel values above the $3\sigma$.
(See text for details.) Error bars indicate the 15\% flux uncertainty in the APEX/LABOCA
maps that dominate the noise distribution. { Dotted lines show the median of the measured
total flux percentage using the two methods.}
}
\label{fig:flux}
\end{figure}

\subsection{Properties of compact sources}\label{sec:cats}

The small-scale structures in the original ATLASGAL maps
have been identified using two complementary methods
in \citet{Contreras2013} and \citet{Csengeri2014}.
In particular, the latter method has been optimised to extract the
properties of the embedded compact sources, and therefore background emission
from the embedding clouds has been systematically removed.
Consequently, adding the large-scale emission does not 
have an impact on the determined compact source properties 
(see also K\"onig et al.\,{ subm}).

\section{Statistical properties of the Galactic cold dust}\label{sec:stat}
\subsection{Dense gas fraction}\label{sec:dgmf}

In Fig.\,\ref{fig:flux} we determine what fraction of the emission is contained in the 
APEX/LABOCA data compared to the  
combined maps in the highest sensitivity part of the \at\ survey, i.e.
in the inner Galaxy with $|\ell |<60^\circ$. 
Since the filtered APEX/LABOCA maps are most sensitive to the
highest column densities, and intrinsically filter the large-scale diffuse gas, 
the flux ratio between the APEX/LABOCA data and the Planck/HFI combined maps is
representative of the dense gas fraction in the cold dust ($f_{\rm DG}$).
In the following, we discuss two methods to estimate $f_{\rm DG}$
based on column densities from the maps and their implications.

First we converted the maps to Jy/pixel units and then
to avoid bias due to varying noise as a function of Galactic longitude, we
applied a $\sigma$-clipping method for both
the APEX/LABOCA only and the combined
maps. 
We determine $\sigma$ for 
each individual tile by fitting a Gaussian function
to the histogram of pixel values considering only the low intensity below 3\,Jy/beam values, and excluding 20\arcmin\ at the edges in Galactic longitude and 15\arcmin\ in Galactic latitude. In this region the flux was then summed up above $3\sigma$.
The ratio of integrated flux in the two maps
gives a Galactic average 
of $f_{\rm DG}\sim5$\%.
Because of the contribution of the lower intensity background
from the Planck/HFI data, in particular, towards the inner Galaxy, 
this method may overestimate the noise in the combined maps, and therefore the 
derived total flux may be underestimated.
This is seen, in particular, in confused regions in the inner Galaxy, where this
method gives higher $f_{\rm DG}$. 
The estimated average $f_{\rm DG}\sim5$\% flux ratio is therefore 
a conservative upper limit
on the dense gas fraction in the cold gas.

We also determine $f_{\rm DG}$ using a more stringent threshold
considering the fraction of dense gas that is likely to be directly involved
in the star formation process. 
Based on Herschel studies in the Gould-belt, \citet{Andre2010} suggests A$_{\rm v}\sim7$ mag ($6.58\times10^{20}$\,cm$^{-2}$)
\footnote{{ We convert H${_2}$ column densities to A$_{\rm v}$ using $N_{\rm{H_2}}/A_{\rm V} = 0.94\times10^{21}$\,cm$^{-2}$\,mag$^{-1}$ \citep{Bohlin1978}.}}
 as a threshold for the formation of dense cores, i.e. the entities 
where star formation takes place (see also \citealp{Lada2010}). 
Using this criterion, we find $f_{\rm DG}<2$\%,
which is a factor of two lower compared to our previous, upper limit estimate.

As shown in Fig.\,\ref{fig:flux}, 
the dense gas fraction is found to be constant 
in the inner Galaxy, despite the large variation in star formation activity. The
only outstanding field is towards the Galactic centre region (the CMZ), 
where both methods give a systematically higher $f_{\rm DG}$ of $10-13$\%.
The higher dense gas fraction found towards the Galactic centre
 is intriguing 
given the low star formation efficiency towards this region (\citealp{Immer2012,U2013mmb,Longmore2013,Csengeri2014, Johnston2014}).

Our estimates provide an independent, and consistent measure of the dense gas
fraction in GMCs, relying on the same tracer for the gas. 
Other commonly used methods rely on comparing observations of dust and molecular 
tracers of gas, or different molecular tracers in external galaxies. 
Despite the difference in the applied method, our results are
in good agreement with the upper limit estimates of $0.07^{+0.13}_{-0.05}$ 
of \citet{Battisti2014}, which are based on comparing
masses of dense clumps derived from the Bolocam Galactic Plane Survey
(BGPS;\,\citealp{Aguirre}),
and GMC masses from the $^{13}$CO (1--0) molecular line data from the Galactic Ring 
Survey~\citep{Jackson2006}. 
For comparison, \citet{Ragan2014} also find  
dense gas mass fractions between 3\% to 18\% for giant filaments, and
based on CO measurements towards more than 90 GMCs identified in the IV$^{th}$ quadrant, \citet{Garcia2014} estimate massive star formation efficiencies
on average of 3\% from their available molecular mass. 
These are in agreement with our results in the present paper. 
Using similar datasets, \citet{Eden2013} estimate $\sim5$\% for 
the clump formation efficiency (CFE), and find no difference
for the CFE between the spiral arm and inter-arm regions. 
Except for the CMZ, 
our results support their conclusions. 
Our findings are also in good agreement with a dense gas fraction between $2-5$\% obtained
for normal spiral galaxies by \citet{Gao2004} using HCN as a tracer for dense gas.

\subsection{Isothermal column density PDFs of the cold Galactic dust}\label{sec:pdf}

\begin{figure}[!htpb]
\centering
\includegraphics[width=6cm, angle=90]{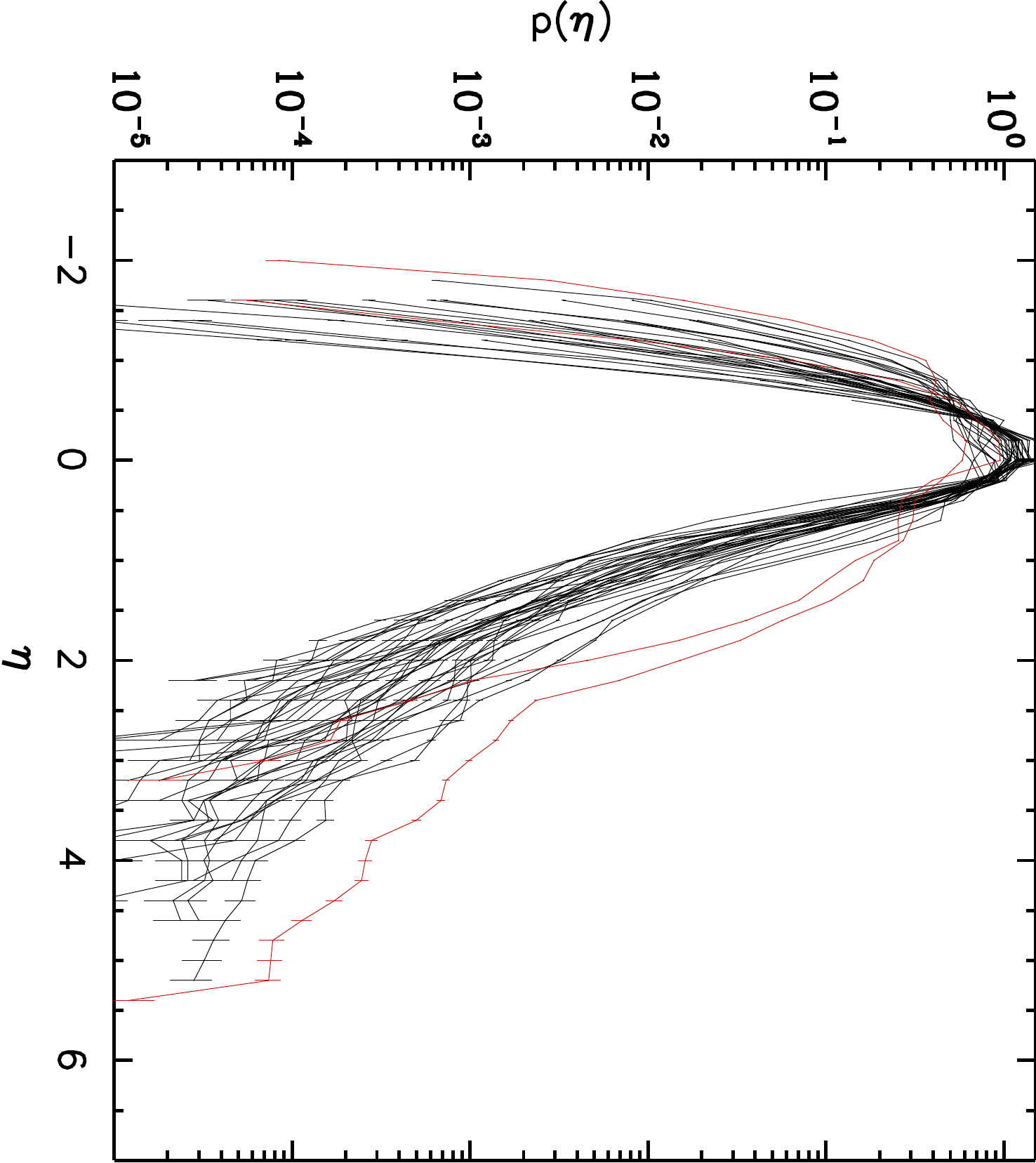}
\includegraphics[width=6cm, angle=90]{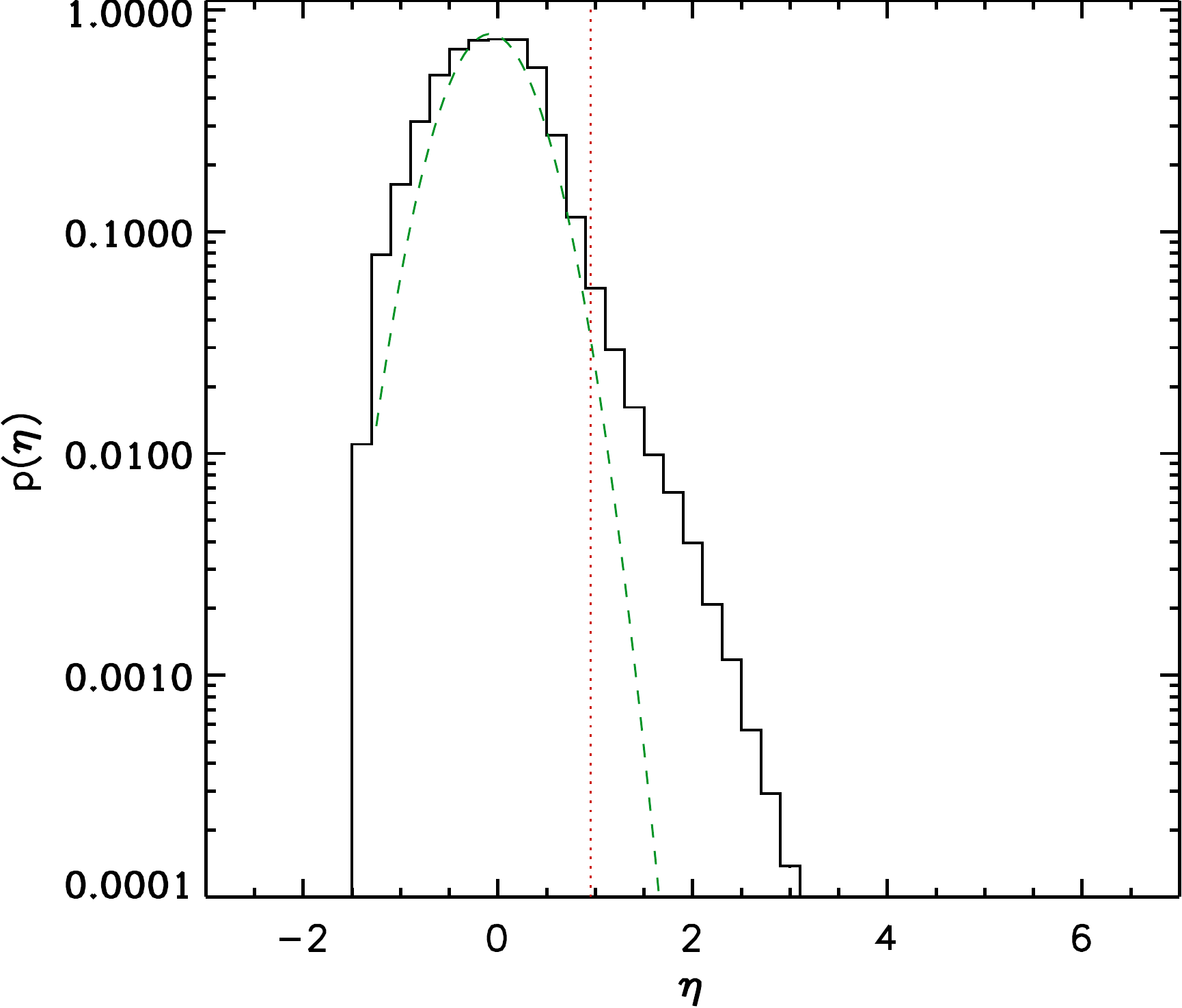}
\caption{{\bf Top:} 
column density probability distribution function (N-PDF) for all 3 degree tiles overplotted.
Only the $|b|$<1$^\circ$ regions are considered due to the increasing
noise towards the edges of the \at\ fields. Red lines show the fields, which include
the Galactic centre.
{\bf Bottom:} The N-PDF of the inner Galaxy ($|\ell|<60^\circ$). Green dotted line shows a
log-normal function with $\sigma=0.41$ at a peak of N$_{\rm H_2}={ 1.10\times10^{22}}$~cm$^{-2}$. Red dotted line shows where 
excess from the power-law tail sets in at $N_{\rm H_2} = { 2.92\times10^{22}}$~cm$^{-2}$.}
\label{fig:pdf}
\end{figure}

To analyse the statistical properties of the structures in the cold Galactic 
dust emission,
we show the H$_2$ column density probability distribution function (N-PDF) for all
fields  (Fig.\,\ref{fig:pdf}, upper panel). 
We calculate the H$_2$ column density for each pixel,
assuming a single dust temperature of $T_{\rm d}=18$~K
and using Eq.\,\ref{eq:nh2}, and we caution that assuming a single dust temperature is
a crude simplification. 
More realistic H$_2$ column density maps should be obtained
using multi-wavelength data from Herschel surveys such as shown 
in e.g.\,\citet{Peretto2010,Schneider2014}.

We define  
$\eta = \rm ln (N_{H_2}$/$<$${\rm N_{H_2}}$$>$) as the natural logarithm of the column
density divided by the mean column density of the corresponding tile. 
{ We only use pixels that are above the lowest $\sim$3$\sigma$
noise level determined from the LABOCA only maps.}
The probability of the
N$_{\rm H_2}$ column density in the range of [N$_{\rm H_2}$,N$_{\rm H_2}+d$N$_{\rm H_2}$] is then given
by $\int^{+\infty}_{-\infty}p (\eta) \rm d\eta = \int_{0}^{+\infty}p ({N_{H_2}}) \rm d{N_{H_2}}=1$;
see also \citet{Schneider2014} for more details.
This formalism follows previous works using numerical simulations to describe the density structure
of molecular clouds \citep{Federrath2008},
and p$_{\eta}$ is adopted to describe a 2D-PDF. 
Projection effects average out the 3D density
distribution and the underlying dense gas fraction of the ISM, and therefore
conversion from 2D column density PDFs to 3D density 
PDFs have to consider the isotropy of molecular clouds. 
Numerical simulations show, however, that systematic differences reflecting the gas 
properties can still be recovered with N-PDFs as well (e.g.\,\citealp{Federrath2015}). 
Conversion to volume densities ($\rho$-PDF) has been discussed in \citet{Kainulainen2014}.

Figure\,\ref{fig:pdf}  (upper panel)  shows the N-PDFs for all 3 by 3 degree tiles. 
This 
representation suffers from line-of-sight contamination (see \citealp{Schneider2014a, Lombardi2015}) 
and does not consider boundaries of molecular clouds. 
The characteristic log-normal shape at low column density is, however, ubiquitous
in the inner Galaxy. Similarly, an excess at high column density regimes
is observed towards all regions, consistent with what was found in  
star-forming regions (e.g.\,\citealp{Kainulainen2009, Andre2010, 
Lombardi2010, Froebrich2010, Schneider2012}).

This excess at high-column densities in the form of a power-law tail 
requires power-law density structures, which 
can be achieved by either a hydrostatic configuration where the power law
arises from a balance of gravitational forces and pressure gradients
(\citealp{Kainulainen2009, Tremblin2014}) or directly in a dynamically
collapsing system \citep{Schneider2013, Schneider2014a}.
 From a theoretical point of view, gravitational collapse
reproduces well the observed high-density excess
(e.g.\,\citealp{Ballesteros-Paredes2011,Kritsuk2011,Federrath2013,Girichidis2014}).
Other explanations, such as non-isothermal flows \citep{Passot1998},
have also been invoked, while magnetic
fields have been shown to slow down the collapse process and reduce
the excess at high densities \citep{Heitsch2001}.

These characteristics show that the APEX/LABOCA and PLANCK/HFI combined maps
trace well a substantial fraction of 
the low-density diffuse material in the Galactic plane, whose line-of-sight 
averaged density structure
resembles that of the more nearby GMCs~
(e.g.\,\citealp{Schneider2012,Russeil2013,Schneider2013}). 

We also point out that the N-PDFs are similar for each field despite their different
properties. The mean of the log-normal fit
to each tile is constant across the Galactic plane and peaks at a value of
N$_{\rm H_2}=1.06\times10^{22}$~cm$^{-2}$. The width of
the log-normal ($\sigma_{\eta}$) shows larger variations with
an average value of $0.41$, and a median of $0.34$. We find the regions covering the CMZ exhibit a more 
significant excess of high-density gas. This is again intriguing because of the low star 
formation efficiency in this region. 

As a reference for extragalactic comparison, 
in Fig.\,\ref{fig:pdf} (lower panel) we show the overall N-PDF describing the
inner $|\ell|<60^\circ$ range of the Galaxy. Again, the low column density part 
of the distribution is well fitted with a log-normal function with
the parameters given above.  
A power-law tail showing an excess at high column densities
is likely associated with star formation (e.g.\,\citealp{Kainulainen2009, Schneider2012}). 
The high-density, power-law tail starts at $\sim$\,${ 2.92\times10^{22}}$~cm$^{-2}$,  and
has a contribution of 2.2\% to the total fraction
of gas, which is consistent with our results obtained in Sect.\,\ref{sec:dgmf},
and corresponds to a Galactic average of dense gas fraction. 
This suggests that the formation of high-density structures is highly inefficient,
and relates to the observed low efficiency of star formation in the Milky Way.
Given the high dense gas fraction towards the CMZ, the formation of high-density
structures, however, cannot be the bottleneck for star formation process to settle in.

\subsection{Mass estimation from dust for the inner Galactic plane}\label{sec:mass}

\begin{figure}[!htpb]
\centering
\includegraphics[width=0.85\linewidth, angle=90]{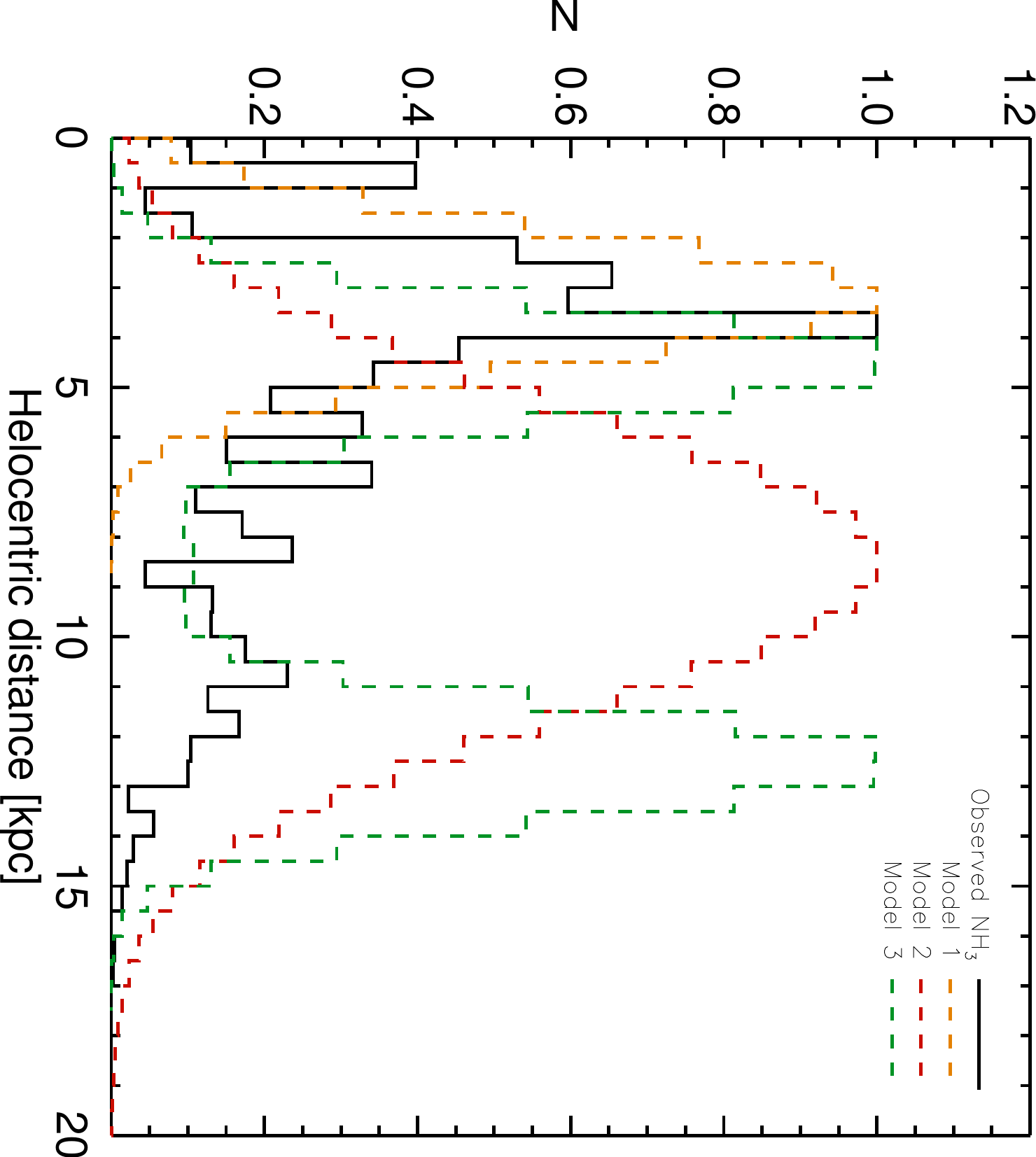}
\includegraphics[width=0.85\linewidth, angle=90]{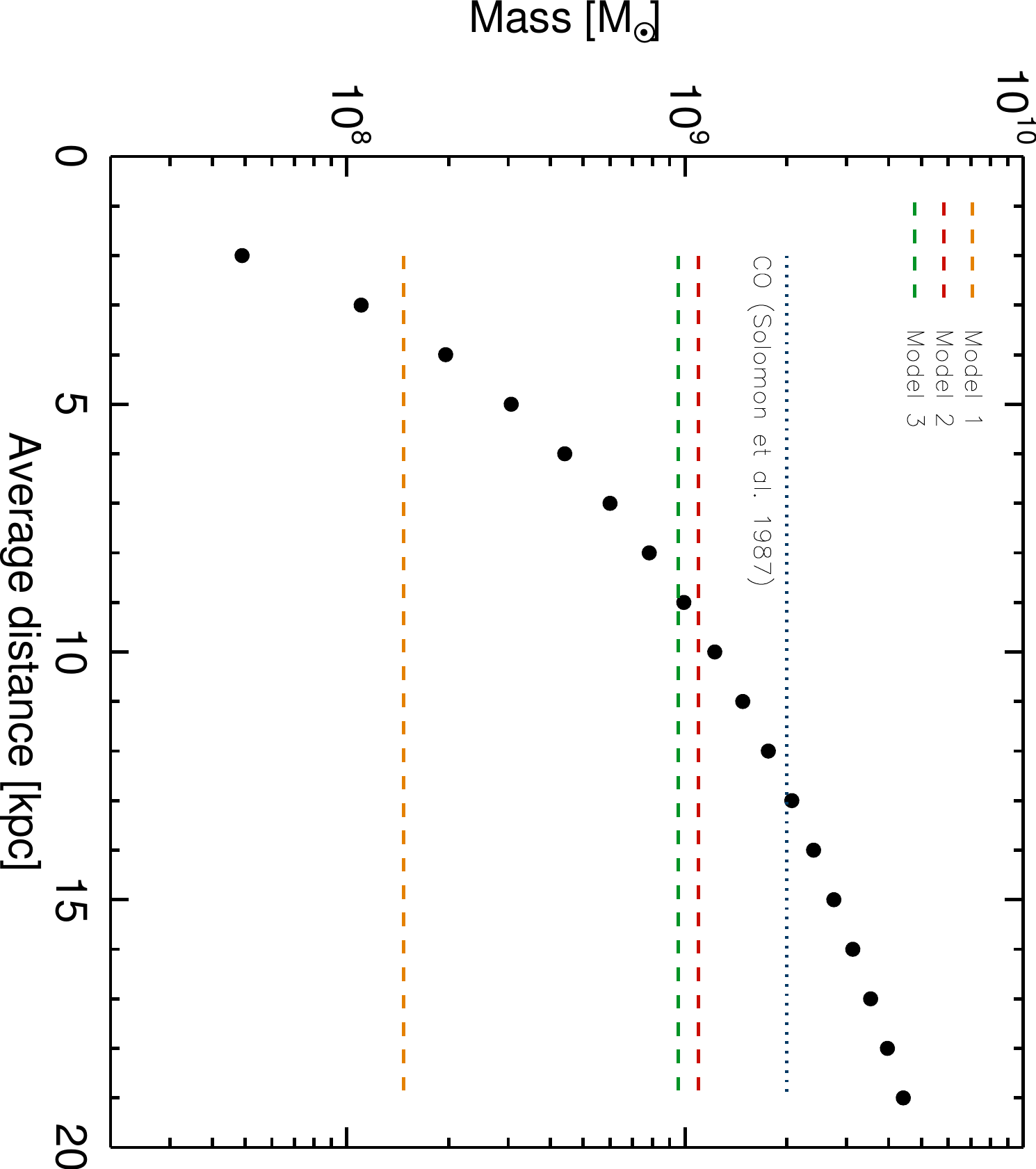}
\caption{
{\bf Top:}
distance distributions used for the mass estimates. Solid line corresponds
to the NH$_3$ observations by \citet{Wienen2012, Wienen2015}.
Dashed lines show the three models discussed in Sect.\,\ref{sec:mass}.
{\bf Bottom:}
molecular gas mass estimate for the inner Galaxy 
($|\ell|<60^\circ$ and $|b|<1.4^\circ$)
from the
ATLASGAL APEX/LABOCA and Planck/HFI combined maps 
of dust emission at 870~$\mu$m. Dots
show the observed total mass as a function of a single adopted distance.
Dashed line shows the mass estimation
estimated with a Monte Carlo simulation.
Red line shows the total estimated
molecular gas mass based on CO measurements~\citep{Solomon1987}.
}
\label{fig:mass}
\end{figure}

Since the ATLASGAL APEX/LABOCA and Planck/HFI combined maps contain emission
at all { large} scales, we use them to assess the global properties of the dust in the
inner Galaxy  to provide a reference for external galaxies. Summing up the total flux
along the Galactic plane, we determine a total flux density of ${ 1.93}$~MJy.
This total flux density is assigned to a mass estimate for the inner Galaxy
in Fig.\,\ref{fig:mass}. Attributing the total emission to a lower and upper distance limit 
of $2$ to $20$~kpc, assuming isothermal dust emission with $T_{\rm d} =18$~K
and the same dust parameters as in Sect.\,\ref{sec:pdf},
we arrive at an upper limit estimate of $M_{\rm tot}\lesssim5\times10^{9}$\,\msol\
as a strict upper limit for the total gas mass in the inner Galaxy.
To obtain a more realistic estimate, we use three models with synthetic distance
distributions using a Monte-Carlo method. In Fig.\,\ref{fig:mass} (top panel)
we show the observed distance distribution from \citet{Wienen2015} with
NH$_3$ measurements towards ATLASGAL sources. In Model-1,
we use a normal distribution of distance estimates adopting the mean 
distance from \citet{Wienen2015}. In Model-2, we use a normal distribution peaking
at the Galactic centre of 8.5 kpc distance \citep{Reid2014}, while in
Model-3 we adopt the sum of three normally distributed components,
one at the Galactic centre, and two symmetric distributions
representing the molecular ring.
We show that despite the different distribution, Model-2 and Model-3
give similar estimates of $\sim$\,${0.87-0.94\times10^9}$\,\msol, while
Model-1 gives a lower estimate of $\sim$\,$0.12\times10^9$\,\msol.
The most recent estimate for the total gas mass in the Milky Way 
is $1.0\pm0.3\times10^9$\,\msol\,\citep{Heyer2015}, with about 70\% of the mass 
within the solar circle.
As a comparison,
the total mass of molecular gas is estimated to be $\sim2\times10^{9}$\,\msol\ 
using CO measurements by \citet{Scoville1975} and \citet{Solomon1987}.
Our estimate based on an independent method is within an order of magnitude
in agreement with these previous studies.

\subsection{Galactic star formation rate}\label{sec:str}
From the total mass estimate and the measure of the dense gas fraction
we can derive the total mass of dense gas in the inner Galaxy.
Using the most realistic model for the distance distribution (Model-3, see Sect.\ref{sec:mass}),
and considering a threshold of N$_{\rm H_2}=6.58\times10^{21}$~cm$^{-2}$
within the smoothed beam of the combined map of 21\arcsec,
we obtain on average $\sim$\,$2\times10^7$\,\msol.
This corresponds to the Galactic average of $\sim$2\% for the fraction of dense gas
and a total mass of $\sim$$10^9$\,\msol.

Direct estimates of star formation rate (SFR) 
in nearby molecular clouds with Herschel \citep{Andre2014}
give a star formation rate per solar mass of $4.5\times10^{-8}$\,yr$^{-1}$,
which is very close to the empirical values derived by \citet{Lada2012}
and \citet{Gao2004}. Based on this we can directly measure a total SFR
in the inner Galaxy of 0.9\,\msol/yr.
Considering the distribution of gas mass, if 70\% of the star formation activity occurs in the inner Galaxy, we obtain a total SFR
of 1.3\,\msol\,yr$^{-1}$ outside the Galactic centre regions.
This is in good agreement with the value of $2.44\pm0.81$\,\msol\,yr$^{-1}$,
estimated by \citet{Csengeri2014} { based on a different, clump and star counting method using} 
the fraction of star-forming versus quiescent
massive clumps in ATLASGAL. Independent estimates of the Galactic SFR
arrive to 2\,\msol\,yr$^{-1}$~\citep{Chomiuk2011}.
Extragalactic studies by e.g.\,
\citet{Wu2005} use a factor of $\sim4$ lower conversion factor for SFR estimates,
which is still consistent with the estimates of Galactic SFR.

A part of the Galactic star formation is also expected in the Galactic centre regions, 
however, these regions exhibit a lower efficiency of star formation 
(e.g.\,\citealp{Csengeri2014}).  While the overall star formation efficiency
is low, the fraction of dense gas is found to be
very high with a fraction up 13\%.
This indicates that the origin of the low star formation efficiency 
is not due to a low efficiency of producing dense gas, but
rather on small scales within the dense gas, which everywhere else in the Galaxy would collapse and form stars at a fast rate.

%

\section{Summary}
We present here reprocessed maps of the ATLASGAL survey, where the APEX/LABOCA
maps have been complemented with the Planck/HFI all-sky survey data at 353~GHz
 to correct for the filtering by ground-based
bolometer observations.
These new maps cover emission at larger angular scales
and, thereby, reveal the structure of cold Galactic dust in great detail.
We summarise our main findings below:

\begin{itemize}
\item[\textbullet]
 A halo of diffuse dust emission is seen towards the direction of 
GMCs. We show two examples, first the $\ell\sim332^\circ$ region where
large structures extending over $0.5^\circ$ seem to bend to the Galactic plane. 
As a second example, the W43 star-forming region is shown with its surrounding
dust halo. We use the full information of dust to estimate H$_2$ column densities
and determine a density contrast of $>5$ between the diffuse halo emission
and the massive ridge of W43 Main from $\sim$\,200 to a few tens of parsec scales.

\item[\textbullet] Adding information from large angular scales helps to better 
   identify the large-scale properties of the cold Galactic interstellar medium.
  Based on visual inspection of the data we show examples of
  giant filaments, which show a linear extension 
  larger than $0.5^\circ$, corresponding to $>30$~pc at distances 
  above 3 kpc.
These filaments have more confined high-density regime
  and similarly to Nessie, they exhibit large
  aspect ratios. 
   
\item[\textbullet]
The presented maps
trace well a substantial fraction of 
the low-density diffuse material in the Galactic plane. 
Using N-PDFs we conclude that the diffuse dust emission shows
very similar properties as a function of Galactic longitude. 
From this, we estimate on average an upper limit on the dense gas fraction
of 5\% on the inner Galaxy and find a relatively constant efficiency
for the formation of dense gas. The Galactic centre region
has a larger dense gas fraction, $f_{\rm DG}\sim13-15$\%, 
which is intriguing considering its low star formation efficiency.
We conclude that the low $f_{DG}$ in the Milky Way is consistent with 
its globally low star formation efficiency. The high $f_{DG}$
observed towards the CMZ suggests that the formation
of dense gas is not the bottleneck for star formation.

\item [\textbullet]
We show that the global N-PDF for the inner Galaxy is
well described by a log-normal distribution, 
which is well explained by the large-scale turbulent nature of the 
ISM.
An excess at high column densities shows 
a power-law tail, which contributes 2.2\% to the total distribution. 
This is a Galactic
average for column densities above ${2.92\times10^{22}}$~cm$^{-2}$. 

\item [\textbullet]
From the total dust emission in the inner Galaxy we provide an independent
estimate of the gas mass of $\sim10^9$\,\msol, which is in good agreement with
the results of CO measurements by \citet{Solomon1987}.

\item [\textbullet]
Based on the mass estimate for the Galactic plane and 
the $f_{\rm DG}$ of 2\%, assuming that 70\% of the star formation
takes place in the inner Galaxy, we estimate a total Galactic
SFR of $\dot M = 1.3$\,\msol\,yr$^{-1}$ outside the Galactic centre regions.

\item [\textbullet]
Although the CMZ is efficient in producing a larger amount of dense gas compared
to other molecular clouds in the Galaxy, the low star formation efficiency suggests that either the local or the global properties of the dense gas must be different from other star-forming regions.
\end{itemize}

With the presented dataset, a more detailed analysis of the dust 
content of the Milky Way is now possible. Furthermore, as also shown by \citet{Schneider2014a},
the presented dataset can be well used to complement the \emph{Herschel} Hi-GAL 
survey \citep{Molinari2010} to derive column density maps of H$_2$
{ at better angular resolution} than only based on  \emph{Herschel}
data.

\begin{acknowledgements}
 This work was partially funded by the ERC Advanced Investigator Grant GLOSTAR (247078) and was partially carried out within the Collaborative Research Council 956, sub-project A6, funded by the Deutsche Forschungsgemeinschaft (DFG).  T.Cs acknowledges support from the \emph{Deut\-sche For\-schungs\-ge\-mein\-schaft, DFG\/}  via the SPP (priority programme) 1573 'Physics of the ISM'.  LB acknowledges support by CONICYT Grant PFB-06.  N.S.  acknowledges supported by the DFG, through project number Os 177/2-1 and 177/2-2 and central funds of the DFG-priority programme 1573 (ISM-SPP).

\end{acknowledgements}

\bibliography{aa.bib}
\bibliographystyle{aa}

\end{document}